\documentstyle[11pt,epsfig]{article}

\newlength{\dinwidth}
\newlength{\dinmargin}
\def\slepton{\widetilde \ell}

\def\squark{\widetilde q}

\def\sneu{\widetilde \nu}
\def\sele{\widetilde e}

\def\sql{\squark_{L}}
\def\sqr{\squark_{R}}

\def\st{\widetilde{t}}
\def\sb{\widetilde{b}}

\def\sz1{{\widetilde{Z}}_{1}}

\def\szk{{\widetilde{Z}}_{k}}
\def\swl{{\widetilde{W}}_{1}}

\def\swi{{\widetilde{W}}_{i}}

\def\mswi{m_{\swi}}

\def\msz1{m_{\sz1}}

\def\mswl{m_{\swl}}

\def\nle{\mathrel{\vcenter
     {\hbox{$<$}\nointerlineskip\hbox{$\sim$}}}}
\def\nge{\mathrel{\vcenter
     {\hbox{$>$}\nointerlineskip\hbox{$\sim$}}}}

\def\stl{\st_{1}}

\def\sth{\st_{2}}
\def\sbl{\sb_{1}}

\def\sbh{\sb_{2}}

\def\mstl{m_{\stl}}
\def\msth{m_{\sth}}

\def\msbl{m_{\sbl}}

\def\mw{m_{W}}
\def\tht{\theta_{t}}
\def\thb{\theta_{b}}

\def\sw{\sin^{2}\theta_{W}}

\def\mtil{\widetilde{m}}

\def\rb{\slash\hspace{-7pt}R}

\def\th{\hat{t}}
\def\sh{\hat{s}}
\begin{document}
~~~\\
\vspace{10mm}
\begin{flushright}
ITP-SU-97/03  \\
hep-ph/9707355 \\
July, 1997
\end{flushright}
\begin{center}
  \begin{Large}
   \begin{bf}
Possible Excess in Charged Current Events 
with High-$Q^2$ at HERA from Stop and Sbottom Production \\
   \end{bf}
  \end{Large}
  \vspace{5mm}
  \begin{large}
Tadashi Kon \\
  \end{large}
Faculty of Engineering, Seikei University, Tokyo 180, Japan \\
kon@ge.seikei.ac.jp\\
 \vspace{3mm}
 \begin{large}
    Takashi Matsushita\\
  \end{large}
Department of Physics, 
Tokyo Metropolitan University, Tokyo 192-03, Japan \\
matsu@heal4.phys.metro-u.ac.jp\\
 \vspace{3mm}
 \begin{large}
    Tetsuro Kobayashi\\
  \end{large}
Faculty of Engineering, 
Fukui Institute of Technology, Fukui 910, Japan \\
koba@ge.seikei.ac.jp\\
  \vspace{5mm}
\end{center}
\vskip50pt
\begin{quotation}
\noindent
\begin{center}
{\bf Abstract}
\end{center}
We investigate a production process 
$e^+p \to \st X \to \sb W^+ X$ at HERA, 
where we consider a decay mode $\sb \to \bar{\nu}_e d$ of the sbottom 
in the framework of an R-parity breaking supersymmetric standard model. 
Both processes of the stop production $e^+ d \to \st$ and 
the sbottom decay $\sb \to \bar{\nu}_e d$ are originated from 
an R-parity breaking superpotential 
$\lambda'_{131} \hat{L}_1 \hat{Q}_3 \hat{D^c}_1$. 
One of signatures of the process should be a large missing transverse 
momentum plus multijet events corresponding to hadronic decays of the $W$. 
It is shown that the signal could appear as an event excess in the 
charged current (CC) processes $e^+p \to \nu X$ with the high $Q^2$ 
at HERA. 
We compare expected event distributions with the CC data recently reported 
by the H1 and ZEUS groups at HERA. 
Methods for extracting the signal from the standard CC processes are 
also discussed. 
\end{quotation}
\vfill\eject
\section{Introduction}

\renewcommand{\thefootnote}{\fnsymbol{footnote}}
Recently reported \cite{H1,ZEUS} event excess at the high $Q^2$ region 
in the deep inelastic (DIS) process $e^+p \to e^+ q X$ 
has very much excited theorists. 
Among its theoretical considerations proposed so far the 
supersymmetric (SUSY) model based approach seems to be an 
appealing possibility. 
It could be interpreted by 
the scalar top (stop) production $e^+p \to \st X \to e^+ d X$ \cite{stoprb} in 
SUSY models with the R-parity breaking (RB) interactions 
proportional to 
$\lambda'_{131}$ \cite{chou,drm,alt,kali,dbst,kim}
\footnote{
Another interesting possibily for "strange stop" scenario, 
$e^+ + s \to {\widetilde{t_{L}}}$ with non-zero coupling $\lambda'_{132}$ 
is discussed in detail by J. Ellis et al. \cite{ellis}.}.  
The excess events seem to be broadly distributed around 200GeV $\sim$ 250GeV in 
the mass ($M$ $=$ $\sqrt{xs}$, where $x$ is the Bjorken parameter). 
In a previous paper \cite{dbst} we have shown that 
it could be simulated by our specific scenario if we 
consider almost degenerate two mass eigenstates $\widetilde{t_1}$ and 
$\widetilde{t_2}$ of the stops with masses $\mstl\simeq 200$GeV and 
$\msth\simeq 230$GeV.

D0 \cite{D0} and CDF \cite{CDF} groups have recently obtained preliminary bounds 
on a leptoquark mass as 
$M_{LQ}$ $>$ 194GeV and 210GeV for Br($LQ \to e^{\pm} q$) = 1.0, respectively. 
These new results suggest that the stop (or leptoquark) with Br($\st \to e^+ d$) = 1.0 
and $m_{\st}$ $\sim$ 200GeV has been excluded. 
Consequently stops could decay into not only $e^+ d$ but also 
$b\swl$ or $\sb W$ via R-parity conserving interactions 
as far as there exists a chargino $\swl$ 
or a scalar bottom (sbottom) $\sb$ lighter than the stops. 
So we can reconcile the interpretation of the high $Q^2$ anomaly at HERA data 
in terms of stop with the Tevatron bound provided that 
Br($\st \to e^+ d$) $=$ $1 -$ Br($\st \to b\swl$ or $\sb W$) $\nle$ 0.8.

Taking into account such situation we are naturally led to examine the processes 
(1) $e^+p \to b  \swl X$ or 
(2) $e^+p \to \sb  W X$.
They will presumably give us unique experimental signatures 
of our scenario. 
We have shown that a possible typical signature of (1) 
would be a high $P_T$ charged lepton plus jet(s) with a large missing 
transverse momentum ${\ooalign{\hfil/\hfil\crcr$P$}}_T$, i.e., 
  one of the signals to be detected at HERA is characterized by the high $P_T$ 
  spectrum of muons \cite{stopsg,matsu}. 
  Note that the lightest neutralino $\widetilde{Z}_1$ and $\sb$ 
possibly decay into $R$-even particles via only non-zero RB coupling 
$\lambda'_{131}$.  
Altarelli et al. have pointed out that another possible signature 
of (1) could be a large ${\ooalign{\hfil/\hfil\crcr$P$}}_T$ plus mutijets, where 
the chargino decays into a neutrino plus quarks \cite{altcc}. 
Moreover, they have shown that such a sigal could be observed as 
an event excess at the high $Q^2$ region in the charged current (CC) DIS, 
$e^+p \to \nu q X$, for a specific sparticle mass spectrum 
$m_{\sele} \nle 70$GeV and $\mswl \simeq 180$GeV.

In the present letter we investigate the process (2) 
mentioned above, i.e., 
we are concerned with a case where the sbottom $\sb$ is lighter than 
the stops and has a mass smaller than about 120GeV. 
Here we assume sufficiently heavy chargino $\swl$ ($\nge 200$GeV) 
and neutralino $\sz1$ ($\nge 100$GeV) for simplicity. 
In this case the stops cannot decay into $b\swl$ and 
the sbottom can only decay into $\bar{\nu}_e d$ via the RB coupling 
$\lambda'_{131}$. 
First we will show that such a light sbottom has not been excluded from the 
present experiments. 
One of characteristic signatures of the process (2) would be 
a large missing transverse momentum plus multijets corresponding to  
hadronic decays of the $W$. It could be observed as 
an event excess at the high $Q^2$ region in the CC DIS. 
We will compare expected event distributions with the CC data recently obtained 
by the H1 \cite{H1,H1trp} and ZEUS \cite{ZEUStrp} groups at HERA.

\section{Model}

We are based on the minimal SUSY standard model 
(MSSM) with an RB superpotential  \cite{Barger};  
\begin{equation}
W_{\rb}=\lambda'_{131}\hat{L}_1 \hat{Q}_3 \hat{D^c}_1, 
\label{RBW}
\end{equation}
where $1$ and $3$ are generation indices. 
We can immediately obtain Yukawa-type interactions from the superpotential (\ref{RBW}) , 
\begin{equation}
L=\lambda'_{131} (
  {\widetilde{t_{L}}} {\overline{d}} P_L e 
+ {\widetilde{e_{L}}} {\overline{d}} P_L t 
+ {\overline{\widetilde{d_{R}}}} {\overline{e^c}} P_L t 
- {\widetilde{b_{L}}} \overline{d} P_L \nu_e
- {\widetilde{\nu_{e}}} \overline{d} P_L b
- {\overline{\widetilde{d_{R}}}} \overline{\nu_e^c} P_L b
) + h.c,  
\label{sqRb}
\end{equation}
where $P_{L}$ reads the left handed chiral projection operator.   
The first term in the Lagrangian (\ref{sqRb}) 
will be most suitable for the $ep$ collider experiments at HERA 
because the stop ${\widetilde{t_{L}}}$ 
will be produced in the $s$-channel for $e^+$-$d$ sub-processes. 
The fourth term will contribute to the sbottom decay, 
${\widetilde{b_{L}}}$ $\to$ ${\bar{\nu}_e}d$. 
Note that the stop ${\widetilde{t_L}}$ 
cannot couple to any neutrinos via the RB interactions. 
That is, no excess is expected in {\it the standard CC events} 
$ep \to \nu q X$ in the squark scenarios. 
As has been pointed out, however, {\it the CC like events} could be expected 
if we consider some appropriate decay chains to the stops (see \cite{altcc} and later).

Here we pay attention to a fact that the stops [sbottoms] 
(${\widetilde{t_L}}$, ${\widetilde{t_R}}$) 
[(${\widetilde{b_L}}$, ${\widetilde{b_R}}$)] are naturally mixed each 
other \cite{stop} due to a large Yukawa coupling of their partner quark 
and the mass eigenstates ($\stl$, $\sth$) [($\sbl$, $\sbh$)] 
are parametrized by a mixing angle $\tht$ [$\thb$], 
\begin{eqnarray}
{\widetilde{t_L}} &=& \stl\cos\tht - \sth\sin\tht, \\
{\widetilde{b_L}} &=& \sbl\cos\thb - \sbh\sin\thb. 
\end{eqnarray}
The interaction Lagrangian (\ref{sqRb}) can easily be rewritten 
in terms of the mass eigenstates. 
In particular, we should note that both stops $\stl$ and $\sth$ could be 
produced by the $eq$ scattering through 
\begin{equation}
L_{\st e d}=\lambda'_{131} (\cos\tht \stl {\overline{d}} P_L e 
                - \sin\tht \sth {\overline{d}} P_L e ) + h.c. . 
\label{stRb}
\end{equation}

The most stringent upper bound on the coupling constant $\lambda'_{131}$ 
comes from the atomic parity violation (APV) experiments 
\cite{Barger,ccu,apv,pdg,napv,apvhera}. 
In the following discussion we adopt rather modest limit, $\lambda'_{131}$ $\nle$ $0.1$. 
We will come back to this point later.

\section{Light sbottom scenario}

Interestingly, in the MSSM there exists a theoretical upper bound on 
the mass of the lighter sbottom $\sbl$. 
Since the left handed stop $\st_L$ and sbottom $\sb_L$ form a $SU(2)$ 
doublet, their masses include the same contribution from a soft scalar breaking mass 
${\mtil_{Q_3}}$, 
\begin{eqnarray}
m^2_{\st_L} &=& {\mtil^2_{Q_3}} + m^2_t + m^2_Z \cos{2\beta} \left( {\frac{1}{2}}-e_u \sw \right) \\
            &=& \cos^2\tht m^2_{\stl} + \sin^2\tht m^2_{\sth}, \\
m^2_{\sb_L} &=& {\mtil^2_{Q_3}} + m^2_b + m^2_Z \cos{2\beta} \left( {-\frac{1}{2}}-e_d \sw \right) \\
            &=& \cos^2\thb m^2_{\sbl} + \sin^2\thb m^2_{\sbh},
\end{eqnarray}
where the last terms in (6) and (8) stand for the $D$-term contributions. 
Combining formulae (6) through (9), we obtain an upper bound on the sbottom mass 
$m^2_{\sbl}$ as, 
\begin{eqnarray}
m^2_{\sbl} &\le& \cos^2\tht m^2_{\stl} + \sin^2\tht m^2_{\sth} - m^2_t + m^2_b - m^2_W \cos{2\beta} \\
           &\le& m^2_{\sth} - m^2_t + m^2_b - m^2_W \cos{2\beta} .  
\label{upper}
\end{eqnarray}
Thus the lighter sbottom $\sbl$ cannot be heavy for relatively light $\sth$. 
For example, $\sbl$ will be lighter than $\stl$ ($\mstl$ $\simeq$ 200GeV) for $m_{\sth}$ 
$\nle$ 250GeV even in the extreme case of $\tan\beta = \infty$ ($\cos{2\beta} = -1$).

Next we examine the decay modes of the stop. 
In the MSSM, the stop could be lighter than the other squarks 
in the first and second generations and the gluino.  
It can decay into various final states : 
$\st$ $\to$ 
 $t\,\szk$ (a),
 $b\,\swi$ (b), 
 $W\,\sb$ (c),  
 $b\,\ell\,\sneu$ (d), 
 $b\,\nu\,\slepton$ (e), 
 $b\,W\,\szk$ (f), 
 $b\,f\,\overline{f}\,\szk$ (g), 
 $c\,\sz1$ (h) and 
 $e\,d$ (i), 
where $\szk$ ($k=1\sim 4$), $\swi$($i=1,2$), $\sneu$ and $\slepton$, 
respectively, 
denote 
the neutralino, the chargino, the sneutrino and the charged slepton. 
(a) $\sim$ (h) are the $R$-parity conserving decay modes, while (i) is 
only realized through the RB couplings (\ref{stRb}). 
If we consider the RB coupling with $\lambda'_{131}$ $>$ $0.01$, 
the decay modes (d) to (h) are negligible due to their large power of 
$\alpha$ arising from 
multiparticle final state or one loop contribution. 
Moreover, in the present case ($\mstl$ $\simeq$ 200GeV) the mode (a) will 
kinematically be suppressed. 
Then only two body decay modes (b), (c) and (i) are left for our purpose.

Here we assume the stops can decay through the modes (c) and (i). 
In other words, we consider sufficiently heavy charginos, 
$\mswi$ $>$ $\msth - m_b$ $\nge$ 220GeV, and a light sbottom, 
$\msbl$ $<$ $\mstl - m_W$ $\nle$ 120GeV. 
The former condition can be easily realized if we take such 
large soft breaking gaugino mass as 
$M_2$ $\nge$ 250GeV (and $|\mu|$ $\nge$ 100GeV). 
As a result of this assumption, the lightest neutralino $\sz1$ becomes 
inevitably 
heavier than about 110GeV because $\msz1$ $\simeq$ $0.5\mswl$ in the MSSM. 
Consequently, an R-parity conserving decay mode of the light sbottom, 
$\sbl \to b\sz1$, is almost kinematically forbidden. 
We can naturally consider $\sbl \to {\bar{\nu}_e}d$ as the dominant decay mode. 
It should be noted that the assumption of relatively light sbottom is 
consistent with the theoretical upper bound on mass (\ref{upper}) for the 
light $\sth$ like $m_{\sth}$ $\simeq$ 230GeV.

Decay widths of $\st_i \to \sb_j W^+$ can be expressed as 
\begin{equation}
\Gamma (\st_i \to \sb_j W^+) = {\frac{g^2}{16\pi}}\left|f_{ij}\right|^2
{\frac{p_j}{m^2_{\st_i}}}\left[m^2_W - 2 \left(m^2_{\st_i}+m^2_{\sb_j}\right) 
+ {\frac{\left(m^2_{\st_i}-m^2_{\sb_j}\right)^2}{m_W^2}} \right],
\end{equation}
where $p_j$ $\equiv$ $\sqrt{E_j^2 - m^2_{\sb_j}}$, 
$E_j$ $\equiv$ $(m^2_{\st_i}+m^2_{\sb_j}-m^2_W)/(2m_{\st_i})$ and 
($f_{11}$, $f_{12}$, $f_{21}$, $f_{22}$) $=$ 
($+\cos\tht\cos\thb$, $-\cos\tht\sin\thb$, $-\sin\tht\cos\thb$, $+\sin\tht\sin\thb$). 
Note that $\Gamma(\st_i \to \sbl W^+)$ is proportional to 
$\cos^2\thb$. 
In Fig.1, we show the branching ratio Br($\stl \to e^+ d$) $=$ 
$1-$ Br($\st \to \sbl W^+$) as a function of $\msbl$, where we take 
$\lambda'_{131}$ $=$ $0.1$, 
$\mstl$ $=$ 210GeV and $\tht$ $=$ 1.0. 
We can find that the desirable branching ratio Br($\stl \to e^+ d$) $\nle$ 0.8 
can be obtained for a wide range of the mass $\msbl$ $\nle$ 120GeV. 
It depends sensitively on the mixing angle $\thb$ through a factor $\cos^2\thb$.

It is natural for us to ask whether or not such a light sbottom, 
$\msbl$ $\nle$ 120GeV, 
has already excluded experimentally. 
In fact, the CDF and D0 groups at Tevatron have excluded 
squarks lighter than about 200GeV \cite{pdg}. 
We should note, however, that the limit has been obtained on the basis of assumptions of 
five degenerate squarks with $m_{\sql}$ $=$ $m_{\sqr}$. 
That is, the expected cross sections will be significantly reduced by a factor ten 
when one sbottom $\sbl$ is lighter than the other squarks. 
At present, the most stringent mass bound on the sbottom with the dominant decay 
mode $\sbl \to {\bar{\nu}_e}d$ is 
\begin{equation}
\msbl \nge 90{\rm GeV}.
\end{equation}
This has been obtained from a stop mass bound $\mstl \nge 90{\rm GeV}$ 
for $\msz1 = 0$ through search for a process, 
$p\bar{p} \to \overline{\stl}\stl X \to c\bar{c}\sz1\sz1 X$ at the D0 \cite{d0stop}. 
Its physics background will be (1) the production processes for the stop and sbottom 
are almost the same at the Tevatron, to which SUSY-QCD diagrams are mainly contributed, 
and 
(2) experimental signatures of the stop and sbottom are the same, i.e., 
large ${\ooalign{\hfil/\hfil\crcr$P$}}_T$ plus jets. 

Next we should examine constraints from the precision measurements at LEP1. 
Potentially, contributions from the $\st_L$ and $\sb_L$ to 
the $\Delta\rho$ could become 
large if their masses are not so different from the weak mass scale of $m_Z$ 
\cite{alt,ellis}. 
We have checked, however, contributions from the $\st_i$ and $\sb_j$ 
to the $\Delta\rho$ 
could become small to such extent as $1\times 10^{-3}$ even for 
$\msbl =100$GeV, $\mstl = 200$GeV and $\msth = 230$GeV if we take large mixing angles 
$\tht \simeq \thb \nge 1.0$ and a large $\tan\beta$ $\nge$ 10. 
Here we have used the formula for $\Delta\rho$ including mixings of both stops and 
the sbottoms \cite{deltarho}. 

\section{Total cross sections}

The analytical expression for the differential cross section of the sub-process 
$e^+ d \to \sbl W^+$ is written as follows, 
\begin{eqnarray*}
&&{\frac{d\hat{\sigma}}{d\th}} = {\frac{g^2{\lambda'_{131}}^2}{128\pi\sh}} 
\Big[\left|{\frac{\cos\tht f_{11}}{D_{\stl}}}
-{\frac{\sin\tht f_{21}}{D_{\sth}}}\right|^2\sh\left[\mw^2-2\left(\sh+\msbl^2\right)
    +{\frac{\left(\sh-\msbl^2\right)^2}{\mw^2}}\right] \\
&&+\left({\frac{\cos\thb}{D_{\nu_e}}}\right)^2
\left(-2\th\left(\sh+\th-\mw^2-\msbl^2\right)
  -2\mw^2\msbl^2+{\frac{\th^2\sh}{\mw^2}}\right) \\
&&-{\frac{2\cos\thb}{D_{\nu_e}}}{\rm Re}\left( {\frac{\cos\tht f_{11}}{D_{\stl}}}
                 - {\frac{\sin\tht f_{21}}{D_{\sth}}}\right)\sh
         \left(-\th+2\msbl^2+{\frac{\th\left(\sh-\msbl^2\right)}{\mw^2}}\right)\Big],
\end{eqnarray*}
where $D_{\st_{1,2}}$ $\equiv$ $\sh-m^2_{\st_{1,2}}+im_{\st_{1,2}}\Gamma_{\st_{1,2}}$ 
and 
$D_{\nu_e}$ $\equiv$ $\th$. 
In this formula, we have included diagrams 
not only for the $\st_{1,2}$ exchanges in the $s$-channel 
but also for the neutrino exchange in the $t$-channel. 

In Fig.2 we show the $\msbl$ dependence of the total cross section 
for the process, 
\begin{equation}
e^+ p \to \sbl W^+ X, 
\label{prc}
\end{equation}
where we take $\mstl$ $=$ 205GeV, $\msth$ $=$ 225GeV and 
$\lambda'_{131}$ $=$ 0.1. 
Two sets of mixing angles (a) ($\tht$, $\thb$) $=$ (0.95, 1.0) and 
(b) (1.2, 1.2) are adopted. They are suitable parameter sets for 
simulation of the excess events at the high $Q^2$ region in the NC DIS. 
Here we use MRS-G parton distribution \cite{MRS}. 
We find that large cross sections of $0.5\sim 1$pb are expected for 
90GeV $\nle$ $\msbl$ $\nle$ 120GeV. 
The double threshold structure originated from both $\stl$ and $\sth$ 
contributions can be seen.

There are two types of experimental signatures of the process (\ref{prc}) ; 
(A) a large missing transverse momentum ${\ooalign{\hfil/\hfil\crcr$P$}}_T$ 
plus multijets and 
(B) 
a high $P_T$ charged lepton ($e^+$ or $\mu^+$) plus mono-jet 
with a large ${\ooalign{\hfil/\hfil\crcr$P$}}_T$. 
(A) and (B) respectively correspond to hadronic decays and 
leptonic decays of the $W^+$ boson. 
In the next section, we will investigate the signal (A), which must be a 
clean signal of the process (\ref{prc}) because of 
Br($W^+ \to q\overline{q'}$) $=$ $2/3$. 
It is worth mentioning that we expect {\it a few events} of the type (B) 
for an integrated luminosity of 30pb$^{-1}$ and a typical efficiency of 
50\% due to $\sigma_{tot} \simeq 1$pb and 
Br($W^+ \to \nu\ell^+$) $=$ $1/9$.

\section{Event excess in CC process}

Now we consider one of signatures for the process, 
$e^+ p \to \sbl W^+ X$, followed by the decays, 
$\sbl \to {\bar{\nu}_e} d$ and $W^+ \to q\overline{q'}$. 
In this case a large ${\ooalign{\hfil/\hfil\crcr$P$}}_T$ plus multijets can 
emerge as its experimental signal. 
Such a sigal could be observed as 
an event excess at the high $Q^2$ region in the CC DIS, because some multijets could not be 
distinguished from the mono-jet even after event selections of the CC DIS. 
In the event selection at HERA, a cut on 
$P_T / E_T$  is used to suppress backgrounds of the 
photoproduction and the NC DIS, where 
$P_T$ $=$ ${\ooalign{\hfil/\hfil\crcr$P$}}_T$ and 
$E_T$ $\equiv$ $\sum_i E_i \sin\theta_i$ (summed over the 
calorimeter cells). 
Indeed, lower cuts on $P_T / E_T$ is efficient in discriminating 
mono-jet events from multijet events because $P_T / E_T$ $\simeq$ 1 
for the former case. 
However, the cuts on $P_T / E_T$ are not so severe in actual analyses 
at HERA, i.e., $P_T / E_T$ $>$ 0.4 and 0.5 at 
the ZEUS \cite{ZEUStrp} and H1 \cite{H1}, respectively. 
Consequently, some mutijet events could be regarded as the CC DIS 
events after the selection. 

We investigate the $Q^2$ and $M$ distributions of the expected events. 
In this analysis, we use the Jacquet-Blondel methods \cite{JB}, 
in which 
$Q_{JB}^2$ and $M_{JB}$ are determined from 
energies and scattering angles of the final jets. 
We combine the 
H1 and ZEUS data explicitly presented in refs.\cite{H1trp,ZEUStrp}. 
Here we use MRS-G parton distribution \cite{MRS} and the hadronization 
has been simulated \`a la JETSET \cite{jetset}.

Shown in Fig.3 are 
$Q^2_{JB}$ and $M_{JB}$ distributions of the expected number of events 
together with the experimental data. 
Figure 3 corresponds to the case that two stops are 
almost degenarate in mass ($m_{\stl}$, $m_{\sth}$) $=$ (205GeV, 225GeV) 
with the finite mixing angle $\tht$ $=$ 0.95. 
Parameters in the sbottom sector are taken as $\msbl$ $=$ 100GeV and 
$\thb$ $=$ 1.0. 
In this case we obtain Br($\stl \to e^+ d$) $\simeq$ 20\% and 
Br($\sth \to e^+ d$) $\simeq$ 10\%. 
That is, the parameter set has not certainly been excluded by the leptoquark 
searches at Tevatron. 
Note, moreover, that the broad mass (or $x$) distribution of the excess events 
in the NC DIS \cite{H1,ZEUS} 
can be reproduced by the mass spectrum and the mixing parameters \cite{dbst}. 
In Fig.3 we take 
$P_T$ $>$ 50GeV and $P_T / E_T$ $>$ 0.5 as kinematical cuts. 
Additional cut, $Q^2$ $>$ 12,500GeV$^2$, is adopted for $M_{JB}$ distribution. 
The tendencies for event excess at the high $Q^2_{JB}$ and 
large $M_{JB}$ of the data 
are successfully reproduced by our scenario 
though in the limited statistics.

We have to emphasize that the signal is characterized by the 
multijet events with a large ${\ooalign{\hfil/\hfil\crcr$P$}}_T$. 
It means that the signal has a broad $P_T / E_T$ distribution. 
Obviously it is different from the one for CC DIS, which has a sharp 
peak at $P_T / E_T$ $\nle$ 1. 
Therefore, we could extract the signal from 
the CC DIS by imposing 
a lower value of cut on $P_T / E_T$ such as $P_T / E_T$ $>$ 0.3. 
Needless to say, detailed jet-analyses will also be indispensable for certainty.

It should be mentioned that the possible event excesses in the 
$Q^2_{JB}$ and $M_{JB}$ distributions are originated from our choice 
of rather large coupling parameter $\lambda'_{131}$ $=$ 0.1. 
If more stringent upper limit, $\lambda'_{131}$ $\nle$ 0.05, 
will be established by elaborated analyses of the APV experiments 
\cite{napv,apvhera}, expected events from the signal would be 
slightly decreased.

\section{Concluding remarks}

We have investigated the process $e^+ p \to \sbl W^+ X$ from the CC channel at HERA. 
Particularly, we considered a case where the sbottom $\sb$ is lighter than 
the stops and has a mass smaller than about 120GeV. 
Here we assume sufficiently heavy chargino $\swl$ ($\nge 200$GeV) 
and neutralino $\sz1$ ($\nge 100$GeV) for simplicity. 
In this case the stops cannot decay into $b\swl$ and 
the sbottom can only decay into $\bar{\nu}_e d$ via the R-parity breaking coupling 
$\lambda'_{131}$. 
We have shown that such a light sbottom has not been excluded by the 
present experiments. One of characteristic signatures of the process could be 
a large missing transverse momentum plus multijets corresponding to  
hadronic decays of the $W$. It could be observed as 
an event excess at the high $Q^2$ region in the CC DIS. 
We have shown that expected event distributions are consistent with 
the CC data recently obtained by the H1 and ZEUS groups at HERA. 
Another signature of our process should be a high $P_T$ charged lepton 
($e^+$ or $\mu^+$) plus mono-jet with a large ${\ooalign{\hfil/\hfil\crcr$P$}}_T$ 
corresponding to leptonic decays of the $W$. 
We can expect {\it a few events} of such a characteristic signal 
assuming the integrated 
luminosity of 30pb$^{-1}$. 
Detailed studies of the leptonic signal of our process are in progress.

\begin{flushleft}
{\Large{\bf Acknowledgements}}
\end{flushleft}
We would like to thank Professors R. Hamatsu, S. Kitamura and M. Kuroda for useful 
discussions. 
One of the present authors (T. Kon) was supported in part by 
the Grant-in-Aid for Scientific Research from the Ministry of Education, 
Science and Culture of Japan, No. 08640388.



\vfill\eject
\begin{figure}
\begin{center}
\mbox{\epsfig{file=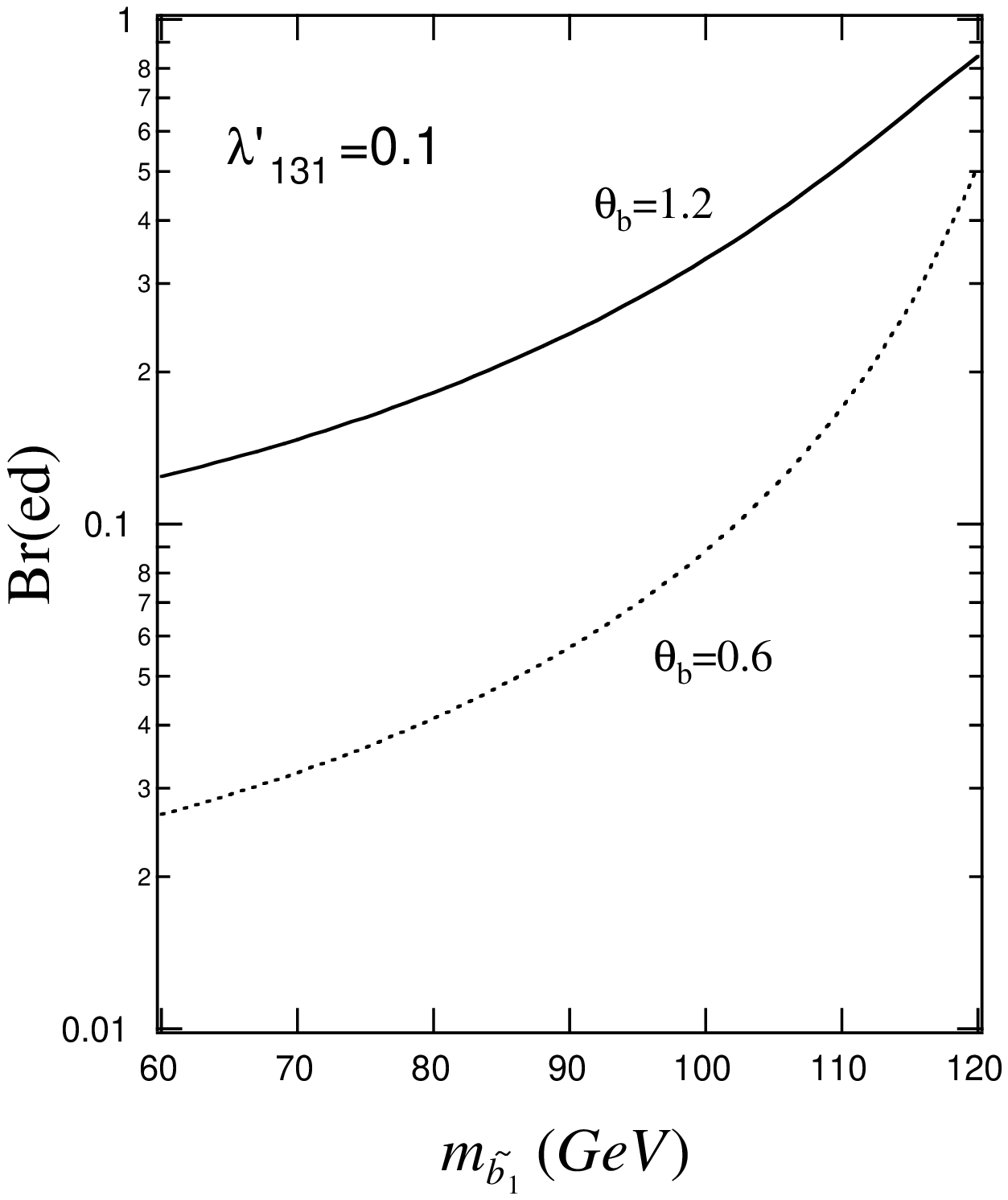,height=0.8\textwidth}}
\caption{
$\msbl$ dependence of branching ratio  
Br($\stl \to e^+ d$) $=$ $1-$ Br($\stl \to \sbl W^+$).
We take $\lambda'_{131}$ $=$ 0.1, 
$\mstl$ $=$ 210GeV and $\tht$ $=$ 1.0. 
}
\end{center}
\end{figure}

\vfill\eject
\begin{figure}
\begin{center}
\mbox{\epsfig{file=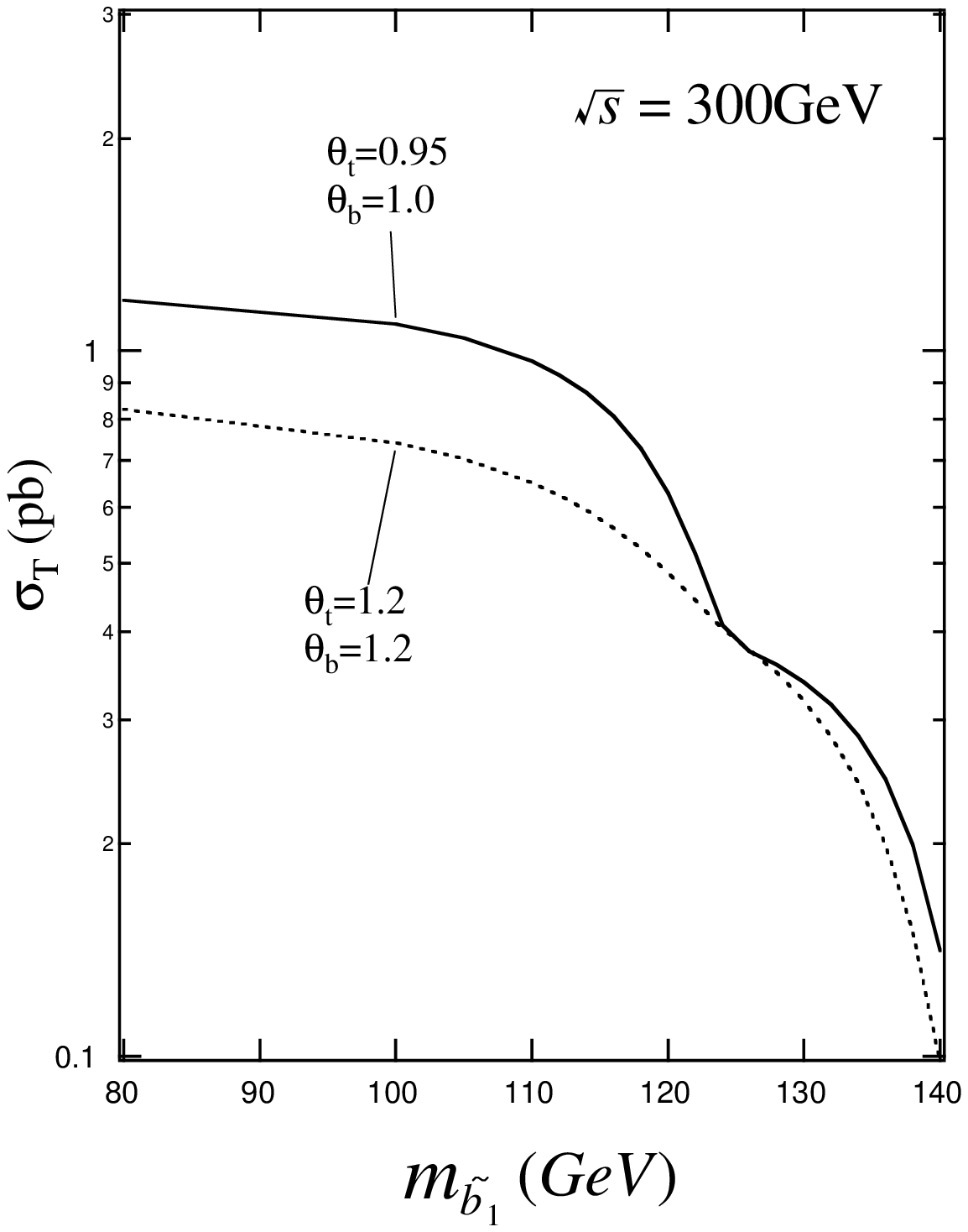,height=0.8\textwidth}}
\caption{
$\msbl$ dependence of the total cross section  
$\sigma (e^+ p \to \sbl W^+ X)$. 
We take 
$\lambda'_{131}$ $=$ 0.1, 
$\mstl$ $=$ 205GeV and $\msth$ $=$ 225GeV. 
}
\end{center}
\end{figure}

\vfill\eject
\begin{figure}
\begin{center}
\mbox{\epsfig{file=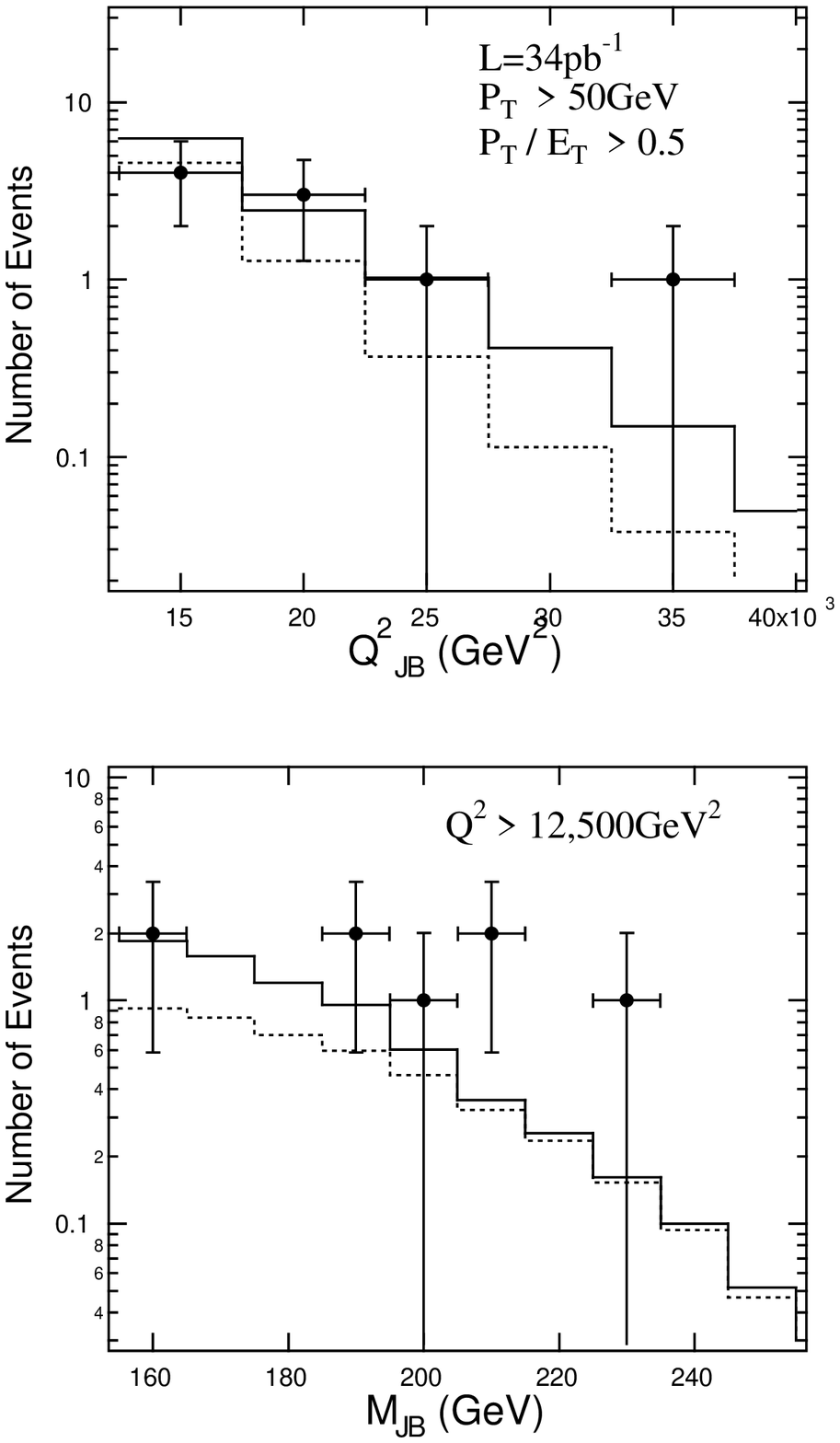,height=0.9\textwidth}}
\caption{
$Q^2_{JB}$ and $M_{JB}$ distribution of the expected number of events 
together with the experimental data.
We take $\mstl$ $=$ 205GeV, $\msth$ $=$ 225GeV, $\msbl$ $=$ 100GeV, 
$\tht$ $=$ 0.95, $\thb$ $=$ 1.0, 
$\lambda'_{131}$ $=$ 0.1 
and integrated luminosity $L=34$pb$^{-1}$. 
As kinematical cuts, $P_T$ $>$ 50GeV and $P_T / E_T$ $>$ 0.5 are adopted. 
Additional cut, $Q^2$ $>$ 12,500GeV$^2$, is taken for $M_{JB}$ distribution. 
Dashed line corresponds to the SM CC expectation. 
}
\end{center}
\end{figure}

\end{document}